# Intertwined Ferroelectricity and Topological State in Two-Dimensional Multilayer

Yan Liang,[1] Ning Mao,[1] Ying Dai,[1,*] Liangzhi Kou,[2,*] Baibiao Huang,[1] Yandong Ma[1,*]

[1]*School of Physics, State Key Laboratory of Crystal Materials, Shandong University, Shandanan Str. 27, Jinan 250100, People's Republic of China*

[2]*School of Mechanical, Medical and Process Engineering, Queensland University of Technology, Garden Point Campus, Brisbane, Queensland 4001, Australia*

*Corresponding author:    daiy60@sina.com (Y.D.); Liangzhi.kou@qut.edu.au (L.K); yandong.ma@sdu.edu.cn (Y.M.)

The intertwined ferroelectricity and band topology will enable the non-volatile control of the topological states, which is of importance for nanoelectrics with low energy costing and high response speed. Nonetheless, the principle to design the novel system is unclear and the feasible approach to achieve the coexistence of two parameter orders is absent. Here, we propose a general paradigm to design 2D ferroelectric topological insulators by sliding topological multilayers on the basis of first-principles calculations. Taking trilayer $Bi_2Te_3$ as a model system, we show that in the van der Waals multilayer based 2D topological insulators, the in-plane and out-of-plane ferroelectricity can be induced through a specific interlayer sliding, to enable the coexistence of ferroelectric and topological orders. The strong coupling of the order parameters renders the topological states sensitive to polarization flip, realizing non-volatile ferroelectric control of topological properties. The revealed design-guideline and ferroelectric-topological coupling not only are useful for the fundamental research of the coupled ferroelectric and topological physics in 2D lattices, but also enable novel applications in nanodevices.

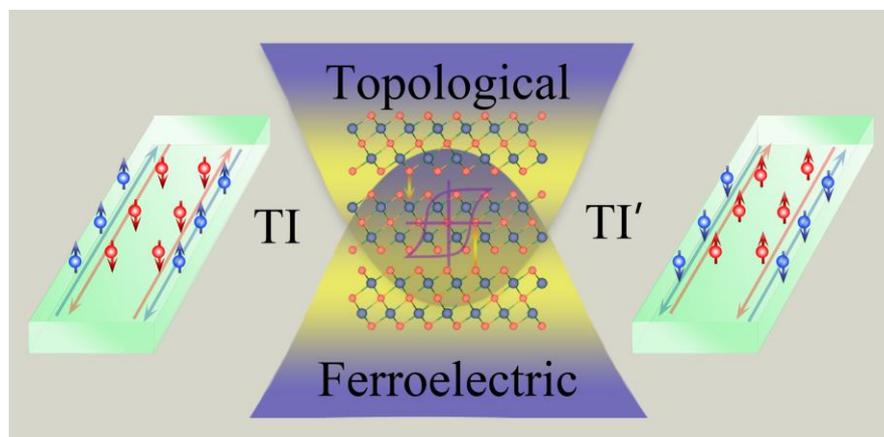

Ferroelectricity and band topology are two intensively investigated yet distinct properties of insulators [1-12]. Physically, there is no inherent exclusion between them due to different origins of polarization and band inversion, their coexistence in a single material leads to the concept of ferroelectric topological insulator (FETI) [3,13-15]. The past years have witnessed the discovery of FETIs, especially in three dimensional, including strained $CsPbI_3$ [16], strained LiZnSb [17], pressured or strained AMgBi [14], and alloyed KMgBi [18], and the induced intriguing electronic properties such as ferroelectric controlled spin vortex. The strongly coupled ferroelectric and topological orders render them both fundamentally intriguing and practically appealing to be used in novel devices.

Unlike 3D FETIs, two-dimensional (2D) lattices with intertwined ferroelectric and topological orders are rather scare [3,19-21], and the coupling of the order parameters are also quite weak in several existing cases [15,22,23]. This is partly due to that, the ferroelectricity in 2D materials has been mainly established in the single-layer asymmetric structures [24,25], while band topology is commonly seen in the materials with heavy elements and strong spin orbital coupling. There, the requirements of symmetrical structure with switchable polarization and band inversion with different parity in the revealed 2D material family have to be simultaneously satisfied, to build the 2D FETIs, which significantly restrict the possible realization of 2D FETIs [9,26-28]. So far, how to expand the scope for material candidates of 2D FETIs, especially with intertwined ferroelectric and topological physics, remains an open question.

Here, based on first-principles calculations, we fill this outstanding gap by introducing a general and simple scheme to realize 2D FETIs with intertwined ferroelectric and topological physics. By studying an example system of trilayer $Bi_2Te_3$, we discover that through a specific interlayer sliding, the charge rearrangement brings the spatial electron-hole separation in van der Waals multilayer based 2D topological insulators. The resultant reversal separation leads to the appearance of both in-plane and out-of-plane ferroelectricity, while the nontrivial topological phase is well reserved, thus enabling the coexistence of ferroelectric and topological orders. The strong coupling between ferroelectric and topological orders is also observed, distinct different topological physics can be induced in such multilayer systems when the ferroelectric polarization is reversed, suggesting the ferroelectric controlled topological properties.

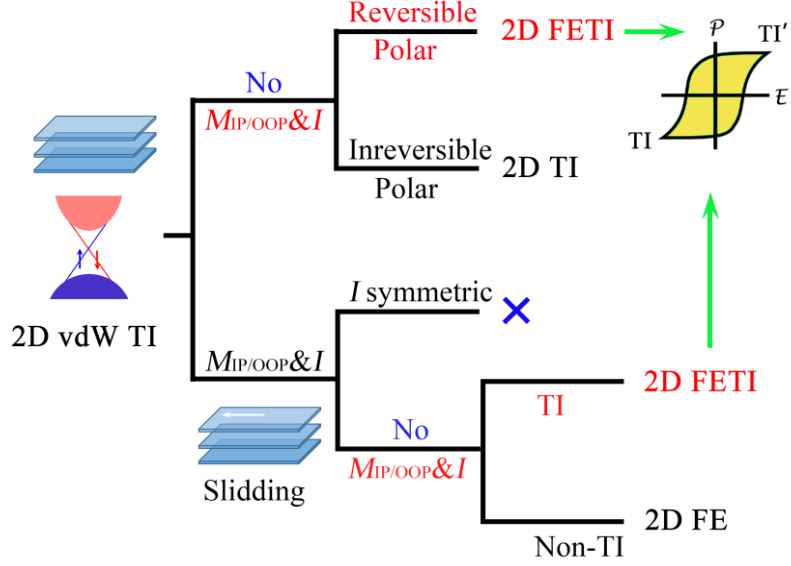

**Figure 1**. (a) Schematic diagram for designing 2D FETIs from 2D van der Waals multilayers with nontrivial topological properties.

The proposed scheme to realize 2D FETIs with intertwined ferroelectric and topological physics is schematically presented in **Figure 1**. Without losing generality, we start from 2D van der Waals multilayers with nontrivial topological properties, the time-reversal symmetry is preserved. Unlike band topology that links with electronic properties [1,4,10,29], ferroelectricity relates to crystal structure symmetry and electric dipole induced by electron distribution [30-34]. To realize ferroelectricity, the polarization has to be switchable. As illustrated in the upper part of **Figure 1**, if the in-plane (IP) and out-of-plane (OPP) mirror symmetries ($M_{IP/OOP}$), as well as the inversion symmetry ($I$), of the 2D multilayer are broken, the ferroelectricity occurs as long as the polarization is switchable, yielding the 2D FETI. The polarization switching is obtained via interlayer translation. If the polarization is unswitchable, it is just a normal 2D TI, without showing ferroelectric order. On the other hand, when the 2D multilayer possesses $M_{IP/OOP}$ or $I$ symmetry, as illustrated in the lower part of **Figure 1**, two different cases can be induced by the interlayer sliding. In first case, the systems like bilayer possess the spatial $I$ symmetry there is no polarization, this configuration is out of our consideration. In second case, the $M_{IP/OOP}$ and $I$ symmetry can be broken by interlayer sliding, the polarization thus appears, and obviously such polarization is electrically switchable. In the latter case, if the topological property is preserved, the ferroelectric-topological phases can be achieved; otherwise, it is a trivial 2D ferroelectric material. As we will show below, the obtained ferroelectric and topological orders in such systems exhibits a strong coupling. This design scheme suggests that the crystal symmetry can be utilized as one screening factor to identify 2D FETIs with intertwined ferroelectric and topological physics.

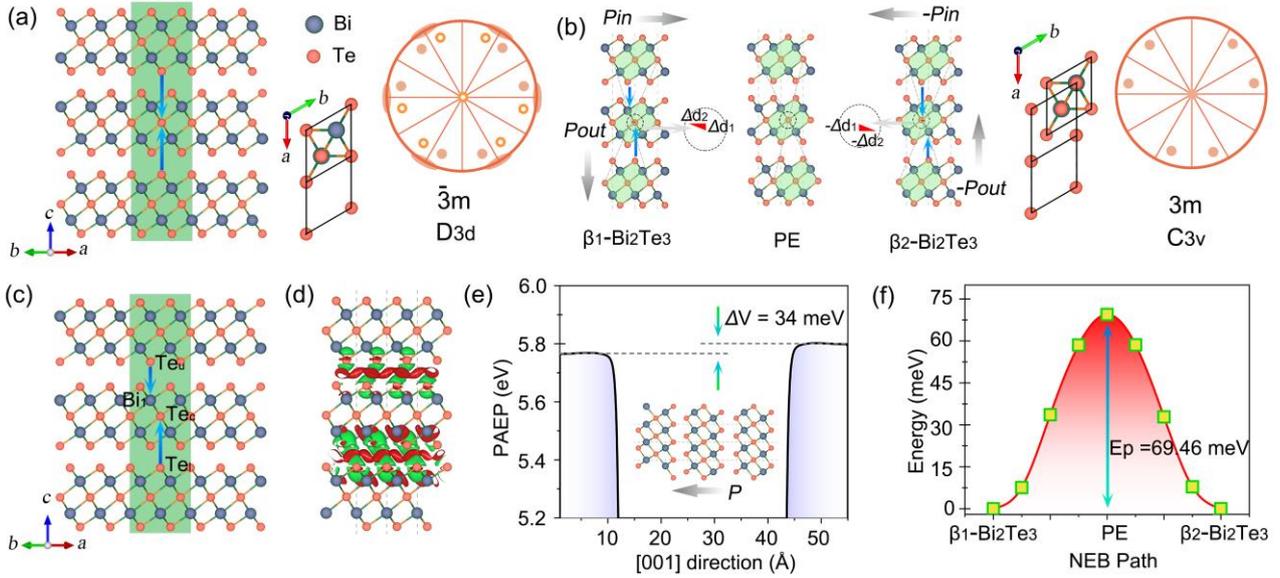

**Figure 2**. Crystal structures of trilayer Bi$_2$Te$_3$ with inversion symmetry (a) and broken inversion symmetry (c). (b) Crystal structures of β$_1$- and β$_2$-Bi$_2$Te$_3$ as well as the schematic diagram of ferroelectric switching. Blue arrows denote the location of interfacial Te atoms of outmost layers. Right parts in (a, b) show the zoomed-in interfacial atomic layers and the equatorial plane projection of point group. (d) Charge density difference and (e) planar average electrostatic potentials along the [001] direction for β$_1$-Bi$_2$Te$_3$. (f) Minimum energy path for ferroelectric polarization reversal between β$_1$- and β$_2$-Bi$_2$Te$_3$.

Following the design scheme, we study the coexistence of FE and band topology in a real material of trilayer Bi$_2$Te$_3$. Our first-principles calculations are performed based on density functional theory as implemented in the Vienna *Ab Initio* Simulation Package (VASP) [35], see Supplemental Material. **Figure 2(a)** shows the crystal structure of trilayer Bi$_2$Te$_3$ (α-Bi$_2$Te$_3$), which are obtained by direct exfoliating from the bulk phase. It shows a space group of $D_{3d}$ with symmetry elements (E, $2C_3$, $3C_2^1$, $i$, $2S_6$, $3\sigma_d$). Clearly, the inversion symmetry prevents it from hosting any polarization. We thus slide the upper and lower quintuple layer (QL) along the [1$\bar{1}$0] and [$\bar{1}$10] directions, respectively, which are referred to as β$_1$- and β$_2$-Bi$_2$Te$_3$, respectively, as shown in **Figure 2(b)**. Such interlayer sliding reduces the space group of trilayer Bi$_2$Te$_3$ to $C_{3v}$ with symmetry elements (E, $3C_3$ and $3\sigma_d$). Due to the simultaneous absence of $I$ and $M_z$ symmetries, β$_1$- and β$_2$-Bi$_2$Te$_3$ host a spontaneous electric polarization of -5.1×10$^9$ e/cm$^2$ and 5.1×10$^9$ e/cm$^2$, respectively, along the out-of-plane (OOP) direction. Obviously, these two polarized configurations can be switched to each other by electric field triggered middle QL sliding [(**Figure 2 (b)**)], and thus correlate to two ferroelectric states, suggesting the OOP ferroelectricity.

To get more insight into the OOP ferroelectricity, we investigate the underlying physics for the electric polarization. In $\beta_1$-Bi$_2$Te$_3$, as displayed in **Figure 2(c)**, the Te$_u$ atom sits above the Bi$_1$ atom, while the Te$_l$ atom lies right below the Te$_c$ atom. The inequivalent distribution of these atoms gives rise to the spatial electron-hole separation along the OOP direction [**Figure 2(d)**], yielding an electric polarization pointing -z direction. The resultant polarization is also suggested by the calculated planar average electrostatic potential of $\beta_1$-Bi$_2$Te$_3$ along the [001] direction. As shown in **Figure 2(e)**, there is a discontinuity ($\Delta V$) of 34 meV between the vacuum levels of the upper and lower QL layers, confirming the formation of electric polarization pointing -z direction. While in $\beta_2$-Bi$_2$Te$_3$, the Te$_u$ atom shifts to above the Te$_T$ atom, while the Te$_l$ atom shifts right below the Bi$_2$ atom; see **Figure 2(b)**. Accordingly, the distribution of these atoms, as well as the spatial electron-hole separation, in $\beta_2$-Bi$_2$Te$_3$ is reversed with respect to that of $\beta_1$-Bi$_2$Te$_3$. Such reversal produces an electric polarization pointing to +z direction for $\beta_2$-Bi$_2$Te$_3$, which are confirmed by the calculated planar average electrostatic potential ($\Delta V$= -34 meV). To evaluate the feasibility of the OOP ferroelectricity in trilayer Bi$_2$Te$_3$, we study the minimum energy path for the ferroelectric switching, which are shown **Figure 2(f)**. The energy barrier is estimated to be 69.46 meV per unit cell, similar which is comparable to the values of other ferroelectrics [33,36-39], indicating its feasibility.

By further examining the distribution of these atoms in the (110) plane [**Figure 2(c)**], we can see that, for $\beta_1$-Bi$_2$Te$_3$, the distance between Te$_u$ and Te$_c$ atoms in the [1$\bar{1}$0] direction is larger than that between Te$_l$ and Te$_c$ atoms. Such imbalance distribution also generates the spatial electron-hole separation along the [1$\bar{1}$0] direction, yielding an in-plane (IP) electric polarization of 3.1×10$^{10}$ e/cm$^2$ pointing to [1$\bar{1}$0] direction. When transforming $\beta_1$-Bi$_2$Te$_3$ into $\beta_2$-Bi$_2$Te$_3$, the Te$_c$ atom moves close to Te$_u$ in the [1$\bar{1}$0] direction. In this regard, the spatial electron-hole separation in the [1$\bar{1}$0] direction is reversed, inducing an IP electric polarization of -3.1×10$^{10}$ e/cm$^2$ pointing [$\bar{1}$10] direction. The reversal of IP electric polarization shares the same energy path as the OOP case. It should be noted, similar to single-layer In$_2$Se$_3$ [40,41], there are three equivalent IP polarizations along the [1$\bar{1}$0], [$\bar{1}$$\bar{2}$0] and [2$\bar{1}$0] directions, which leads to a zero net IP polarization. When introducing the substrate effect, the three-fold rotation symmetry is broken to realize the IP ferroelectricity, which has been well demonstrated in experiment [34,42-44]. Accordingly, both IP and OOP ferroelectricity can be realized in trilayer Bi$_2$Te$_3$.

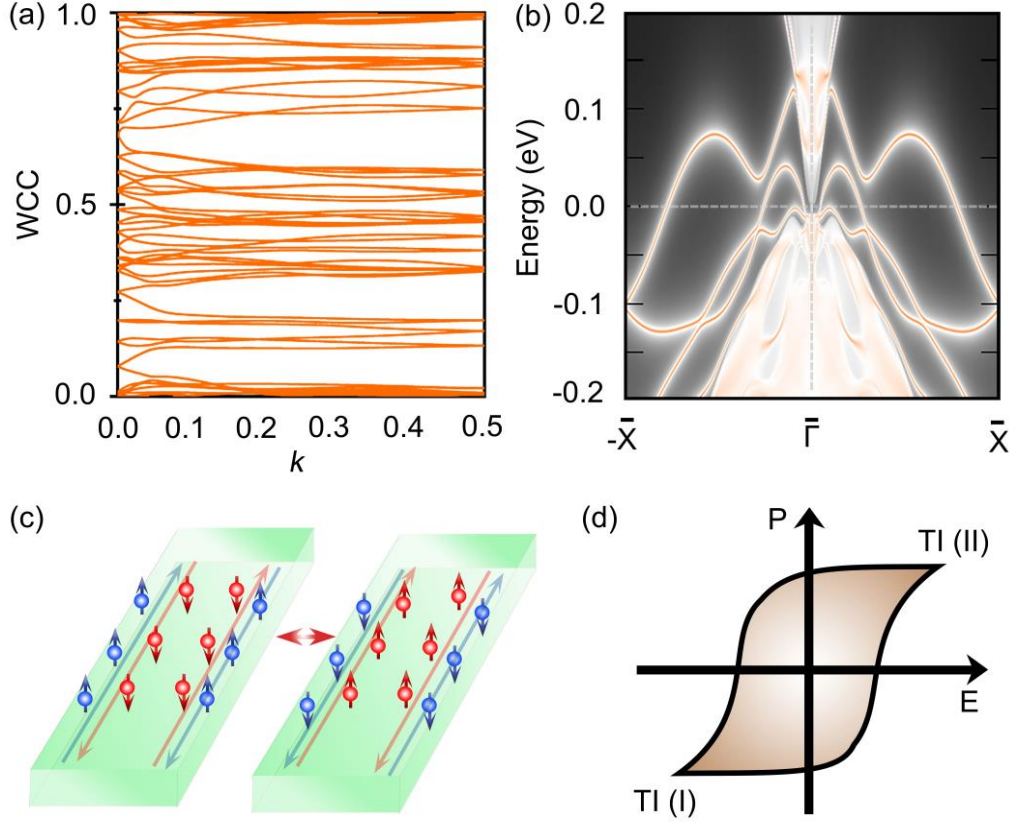

**Figure 3**. (a) Evolution of the WCC along $k_y$ for $\beta_1$-$Bi_2Te_3$. (b) TB band structure of armchair-edged nanoribbon of $\beta_1$-$Bi_2Te_3$ obtained by MLWFs, showing edge states inside the gap of bulk bands. (c) Proposed electrical switch of the chirality and direction of the spin-locked current at boundary. (d) Schematic representation of 2D FETI with two polar TI states.

Next, we study the electronic properties of trilayer $Bi_2Te_3$ in the ferroelectric phase. As $\beta_1$- and $\beta_2$-$Bi_2Te_3$ are linked as two equivalent ferroelectric states, here we take $\beta_1$-$Bi_2Te_3$ as an example. **Figure S1(a)** shows the band structure of $\beta_1$-$Bi_2Te_3$ without including spin-orbit coupling (SOC), from which we see that it is an indirect gap semiconductor with a global gap of 0.51 eV near the $\Gamma$ point. By analyzing orbital contributions, we find the highest valence bands (VB) near the Fermi level is mainly contributed by Te-$p$ orbital, while Bi-$p$ orbital makes the dominate contribution to the lowest conduction bands (CB). Upon taking SOC into account, the VB and CB bands near the $\Gamma$ point experience a significant Rashba spin splitting [**Figure S1(b)**], which can be attributed to the existence of OOP electric polarization in $\beta_1$-$Bi_2Te_3$. The corresponding Rashba parameter is calculated to be $\alpha_R = \frac{2E_R}{k_0} = 0.67$ eVÅ. When SOC effect is considered, it is interesting to notice that the CBM and VBM move closer and the band gap is reduced to 9 meV. Such band narrowing and M-shaped VBM normally indicates a nontrivial topological phase.

To confirm the nontrivial topological order in $\beta_1$-Bi$_2$Te$_3$, we calculate the topological invariant $Z_2$. Due to its broken inversion symmetry, the $Z_2$ invariant is calculated by tracing the Wannier charge center (WCC) using non-Abelian Berry connection [45]. The Wannier functions (WFs) related with lattice vector R can be written as:

$$|u_{nk}> = \frac{1}{2\pi}\int_{-\pi}^{\pi} dk e^{-ik(R-x)}|u_{nk}>$$

Here, a WCC is defined by the mean value of $<0_n|\hat{X}|0_n>$, where the $\hat{X}$ represent the position operator and $0_n$ is the state corresponding to a WF in the cell with R = 0. Then we obtain:

$$\bar{x}_n = \frac{i}{2\pi}\int_{-\pi}^{\pi} dk <u_{nk}|\nabla_k|u_{nk}>$$

Assuming $\sum_\alpha \bar{x}_\alpha^S = \frac{1}{2\pi}\int_{BZ} A^S$ with S = I or II, the summation in α is the occupied states and A is Berry connection. So we get the $Z_2$ invariant following

$$Z_2 = \sum_\alpha [\bar{x}_\alpha^I\left(\frac{T}{2}\right) - \bar{x}_\alpha^{II}\left(\frac{T}{2}\right)] - \sum_\alpha [\bar{x}_\alpha^I(0) - \bar{x}_\alpha^{II}(0)]$$

The calculated evolution of WCC is shown in **Figure 3(a)**. As expected, the WCC is crossed by any arbitrary horizontal reference lines an odd number of times, indicating $Z_2$ = 1. This firmly confirms the nontrivial topological phase of $\beta_1$-Bi$_2$Te$_3$. As the existence of the localized metallic helical edge channels is the prominent feature for 2D TI, we calculate the armchair edge states by using a tight-binding (TB) Hamiltonian in the maximally localized WF. As shown in **Figure 3(b)**, a pair of edge states around the edge projected Γ-point are observed within the bulk gap. And these states are robust and spin helical, where opposite spin polarizations are propagated along the different directions. The topological edge states further manifest the nontrivial properties. As a result, the coexistence of ferroelectric and topological orders is obtained in trilayer Bi$_2$Te$_3$.

In the following, we discuss their coupling of ferroelectricity and topological orders in trilayer Bi$_2$Te$_3$. Different from 2D TI with inversion symmetry, due to the existence of IP electric polarization, the characters of the nontrivial edge states contributed by two opposite zigzag edges would be remarkably anisotropic. Taking the edge states along [1$\bar{1}$0] and [$\bar{1}$10] as examples, we show them in **Figure S2**. As expected, these two edge states are distinctly different. Under the ferroelectric switching, these two different nontrivial edge states would be exchanged. This means that the character of nontrivial edge state, such as the position of Dirac point, at an assigned edge can be precisely controlled by ferroelectricity. Moreover, because of the coupling between IP and OOP ferroelectricity, either IP or OOP external electric field can trigger such modulation. This results in the coupled ferroelectric and topological physics in such multilayer systems. Benefit from such ferroelectric-topological coupling, the fascinating topological *p-n* junctions can be easily obtained when forming a side-by-side ferroelectric domain walls with opposite polarizations [46-48]. In addition, utilizing the either IP or OOP external electric field, such topological *p-n* junctions are controllable.

Meanwhile, for such multilayer, the two ferroelectric states with opposite polarizations are linked to each other through an inversion operation. As a result, as shown in **Figure 3(c)**, the chirality as well as the direction of the spin-locked currents at boundaries are closely associated with the direction of ferroelectric polarization. In other words, the direction of topological spin current can be fully controlled by ferroelectricity, which would promote novel applications in conceptually new devices. Moreover, due to the direction of spin-locked current can be viewed as a ferroic order, such multilayers can also be treated as multiferroic systems, see **Figure 3(d)**, holding potential for novel multiferroic devices. We wish to stress that these coupled ferroelectric and topological physics are not limited to trilayer $Bi_2Te_3$, but applicable for all 2D FETIs designed by this scheme.

In summary, we introduce a general scheme to realize coupled ferroelectric and topological physics in multilayer systems. Taking trilayer $Bi_2Te_3$ as an example, we show that through a specific interlayer sliding, both IP and OOP ferroelectricity can be realized in van der Waals multilayer based 2D topological insulators, resulting in the coexistence of ferroelectric and topological orders. We further show that the ferroelectric and topological orders exhibit a strongly coupling. Under the ferroelectric switching, distinct different topological physics can be induced in such multilayer systems.

**ACKNOWLEDGMENTS**

This work is supported by the National Natural Science Foundation of China (No. 11804190, 12074217), Shandong Provincial Natural Science Foundation (Nos. ZR2019QA011 and ZR2019MEM013), Shandong Provincial Key Research and Development Program (Major Scientific and Technological Innovation Project) (No. 2019JZZY010302), Shandong Provincial Key Research and Development Program (No. 2019RKE27004), Shandong Provincial Science Foundation for Excellent Young Scholars (No. ZR2020YQ04), Qilu Young Scholar Program of Shandong University, and Taishan Scholar Program of Shandong Province.